\documentclass[paper]{aa}
\usepackage{txfonts}
\usepackage{natbib}
\usepackage{graphicx}
\bibpunct{(}{)}{;}{a}{}{,}
\begin{document}

\title{The abundance of HNCO and its use as a diagnostic of environment}
\author{D~M. Tideswell \inst{1} \and G.~A. Fuller
  \inst{1}\and T.~J. Millar \inst{2} \and A.~J. Markwick \inst{1}}
\institute{Jodrell Bank Centre for Astrophysics, Alan Turing Building,
  School of Physics and Astronomy, The University of Manchester,
  Oxford Road, Manchester M13 9PL, UK. \and Astrophysics Research
  Centre, School of Mathematics and Physics, Queen's University
  Belfast, Belfast, BT7 1NN, UK.}
\date{Received ** / Accepted **} 

\abstract{}{We aim to investigate the chemistry and gas phase
  abundance of HNCO and the variation of the HNCO/CS abundance ratio
  as a diagnostic of the physics and chemistry in regions of massive
  star formation.}  {A numerical-chemical model has been developed
  which self-consistently follows the chemical evolution of a hot
  core.  The model comprises of two distinct stages. The first stage
  follows the isothermal, modified free-fall collapse of a molecular
  dark cloud. This is immediately followed by an increase in
  temperature which represents the switch on of a central massive star
  and the subsequent evolution of the chemistry in a hot, dense gas
  cloud (the hot core).  During the collapse phase, gas species are
  allowed to accrete on to grain surfaces where they can participate
  in further reactions.  During the hot core phase surface species
  thermally desorb back in to the ambient gas and further chemical
  evolution takes place. For comparison, the chemical network was also
  used to model a simple dark cloud and photodissociation regions.}
{Our investigation reveals that HNCO is inefficiently formed when only
  gas-phase formation pathways are considered in the chemical network
  with reaction rates consistent with existing laboratory data. This
  is particularly true at low temperatures but also in regions with
  temperatures up to $\sim$200~K.  Using currently measured gas phase
  reaction rates, obtaining the observed HNCO abundances requires its
  formation on grain surfaces - similar to other `hot core' species
  such as CH$_3$OH.  However our model shows that the gas phase HNCO
  in hot cores is not a simple direct product of the evaporation of
  grain mantles.  We also show that the HNCO/CS abundance ratio varies
  as a function of time in hot cores and can match the range of values
  observed.  This ratio is not unambiguously related to the ambient UV
  field as been suggested - our results are inconsistent with the
  hypothesis of Mart\'in et al (2008).  In addition, our results show
  that this ratio is extremely sensitive to the initial sulphur
  abundance. We find that the ratio grows monotonically with time with
  an absolute value which scales approximately linearly with the S
  abundance at early times.}{}

\maketitle

% introduction to HNCO/hotcores/modelling & obs history

\section{Introduction}

The molecule HNCO (isocyanic acid) was first detected in the
interstellar medium over 30 years ago \citep{sb1972} in the Sgr B2
molecular cloud complex where its distribution was found to be
spatially extended with relatively strong emission.  Since its
discovery, HNCO has been subsequently observed in several other
molecular clouds, including the dark cloud TMC-1
\citep[e.g.][]{b1981,getal1982, jab1984}.  In the survey of 18
molecular clouds conducted by \citet{jab1984}, HNCO emission was
detected in seven sources with an average excitation temperature of
12~K, suggesting that this molecule was tracing cold
gas. \citet{jab1984} proposed that HNCO was a dense gas tracer due to
the coincidence of the emission with regions of high density
($n\gtrsim10^{6}\ \mathrm{cm^{-3}}$).  A larger survey of 81 molecular
clouds was conducted by \citet{zhm2000}.  They reported a 70\%
detection rate with typical fractional abundances relative to the
total hydrogen number density, $n_{\mathrm{H}}$, of $\sim10^{-9}$ and
a wide range of rotational temperatures from 10~K to 500~K.  Three of
the objects were mapped and showed the emission region to be compact
and centrally peaked.

Isocyanic acid has also been detected in hot cores (HCs), the sites of early
stage, high mass star-formation including G34.3+0.15 \citep{metal1996},
W3(H$_2$O) \citep{hv1997} and Sgr B2; in the N,M and NW cores
\citep{netal2000}.  A survey towards seven high-mass star forming regions by
\citet{betal2007} identified HNCO with fractional abundances relative to
$n_{\mathrm{H}}$ of between 6.4$\times10^{-9}$ and 5.4$\times10^{-8}$ and
excitation temperatures from 64~K to 278~K.  The observations suggested that
HNCO can also trace much warmer gas - consistent with the findings of
\citet{zhm2000} and as such could be a potential indicator of star formation
activity.

A recent multitransition study of 13 molecular clouds towards the
Galactic Centre was conducted by \citet{metal2008}.  The sources were
selected to represent a range of physical conditions and included
photodissociation regions (PDRs) along with HCs and giant molecular
clouds. \citet{metal2008} compared the CS abundance, a species used as
a dense gas tracer and is regarded as a tracer of star formation sites
\citep{betal1996}, to that of HNCO for each of the sources.  They
found that the ratio of the abundance of HNCO to that of CS (as traced
by $^{13}$CS) was smaller for regions in which the FUV radiation is
enhanced (PDRs), perhaps indicating that HNCO is more sensitive than
CS to FUV photodissociation. The HNCO/CS ratio was systematically
smaller in PDR-like regions by approximately two orders of magnitude
than in the giant clouds. HCs were found to have intermediate
values. It was therefore proposed that the HNCO/CS abundance ratio may
provide a useful tool in distinguishing between shock and radiation
(FUV) activity in molecular clouds.  However, the \citet{metal2008}
survey only observed a limited sample towards the Galactic Centre and
before the usefulness of this possible tracer can be established more
complete surveys and a theoretical understanding of the processes
involved are desirable.

Since the detection of HNCO there has been much speculation as to how
this molecule forms but there have been few chemical models developed
which incorporate an HNCO reaction network.  Some authors have
considered formation only in the gas phase
\citep[e.g.][]{tth1999,i1977} whilst others have modelled the
molecule's formation on grain surfaces \citep[e.g.][]{gwh2008}.  Here
we reconsider these processes and motivated by the observations of
\citet{metal2008}, construct a model for a hot core including both gas
phase and grain surface formation and destruction of HNCO.  We set out
to model both the HNCO abundance and the HNCO/CS ratio as a function
of time for such a region.  We also aim to explore the response of the
gas-phase formation of HNCO to changing physical conditions.  Besides
modelling HNCO in hot cores, simple models of a dark cloud and PDRs
are also presented for comparison.

\section{Model descriptions}
\label{model}
Our hot core model consists of two distinct stages. The first stage is
the collapsing cloud (CC) phase which follows the isothermal collapse
of a dark molecular cloud whilst the second stage follows the
evolution at constant density and a specified temperature
%subsequent heating of the collapsed cloud resulting from the formation
%of a central massive star 
- this is the HC phase.  Along with the CC
and HC models simple dark cloud (DC) and PDR models were also
produced.  These models were similar to the HC models in that no
collapse was included but different physical conditions were used.
All the models described in this paper were single point models.

\subsection{Physical model}

The CC model begins at a time $t=0$ with an homogeneous dark molecular
cloud at a temperature, $T$ and an initial total hydrogen number
density of 2$\times10^4\ \mathrm{cm^{-3}}$.  The total hydrogen number
density is defined as $n_\mathrm{H} = 2 n(\mathrm{H_2})+n(\mathrm{H})$
and is equal to the cloud's density.  It was assumed that the cloud
was shielded by a visual extinction, $A_\mathrm{v}$, of 15 magnitudes
and the standard Galactic UV radiation field and cosmic ray ionisation
rates were adopted.  These physical conditions were chosen to be
representative of a typical Galactic molecular cloud
\citep[e.g.][]{mh1990}.  The cloud was then allowed to collapse
isothermally following the modified free-fall collapse formula
\citep[see][]{spitzer1978};
\begin{equation}
\label{modifiedcollapse}
\frac{\textrm{d}n(t)}{\textrm{d}t} = B \left( \frac{n(t)^4}{n_{\mathrm{o}}} \right)^{1/3} \left\{ 24 \pi G m_{\mathrm{H}} n_{\mathrm{o}} \left[ \left( \frac{n(t)}{n_{\mathrm{o}}} \right)^{1/3} - 1 \right] \right\}^{1/2}~\textrm{cm}^{-3} \textrm{s}^{-1},
\end{equation}
where $B$ is the retardation factor and a value of 0.7 was adopted
following \citet{retal1992}, $n_\mathrm{o}$ is the initial cloud
density (total hydrogen number density) prior to the collapse being
initiated, $G$ is the Universal Gravitational constant and
$m_{\mathrm{H}}$ is the atomic mass of hydrogen.

The collapse was halted once $n_\mathrm{H}=2\times10^7\
\mathrm{cm^{-3}}$ which occurs after approximately 5$\times10^5$
years.  This density was chosen to be representative of a typical hot
core.  As soon as this density was reached the model terminated and
this was considered to be the end of the CC stage.  The final chemical
states of these collapse models were used as the initial chemistry of
the subsequent hot core phases. A summary of the physical parameters
adopted for the seven CC models used in this work is given in Table
\ref{ccmodels}.

\begin{table}
\centering
\caption{Summary of the seven collapsing cloud models. $T$ is the isothermal temperature of the collapse. }
\label{ccmodels}
\begin{tabular}{ccl}
\hline\hline
Model & $T$ (K) & HNCO formation chemistry\\
\hline
CC1 & 10 & All\\
CC2 & 10 & Gas phase only\\
CC3 & 10 & Surface only (reactions a,b)\\
CC4 & 10 & All, reduced sulphur\\
CC5 & 20 & All\\
CC6 & 50 & All\\
CC7 & 100 & All \\
\hline\hline
\end{tabular}
\end{table}

The HC model represents the second stage of the overall 
% hot core
evolution.  The HC was considered to be an homogeneous cloud at the
final density reached at the end of the collapse phase,
$n_\mathrm{H}=2\times10^7\ \mathrm{cm^{-3}}$, in which the
$A_\mathrm{v}$ was high enough that the influence of external FUV
photons is negligible.  For the duration of the HC model these
physical conditions remained fixed.  
%The start of the HC model was
%defined by an instantaneous increase in the ambient temperature.

A total of 10 HC models were produced.  Models 1 through 4 all had the
same physical conditions including a temperature of 200~K, meant to
represent the heating by a newly formed central star. The four models
differ in their chemical networks.  On the other hand, models 5 to 10
all had the same chemistry but differing temperatures (the temperature
of the initial the collapsing clouds for these models also differed).
A summary of these models can be seen in Table \ref{models_used}

In order to follow chemistry on dust surfaces it was necessary to
include dust grains in the models. By assuming a gas mass to dust mass
ratio of 100, the dust number density (relative to $n_\mathrm{H}$) was
calculated to be 1.33$\times10^{-12}$. A single population of
spherical dust grains was then assumed with each grain having a radius
of $10^{-5}$ cm.  The total number of binding sites on each grain was approximately $10^6$ assuming a surface density of 7.9$\times10^{14}\ \mathrm{cm^{-2}}$.

The DC and PDR models were modelled in a similar fashion to the HC. The
physical conditions were kept constant in each case and no collapse was
included. The DC was modelled using a $T=10$~K, $n_\mathrm{H}=2\times10^4\
\mathrm{cm^{-3}}$ and an $A_\mathrm{v}=15$, whilst the PDR model adopted
conditions representative of the Horsehead nebula \citep{hetal2005}; $T=50$~K,
$n_\mathrm{H}=2\times10^4\ \mathrm{cm^{-3}}$, $A_\mathrm{v}$ $=$ 1 and an
enhanced FUV radiation field equal to 60 times the interstellar value as used
in all the other models.  In order to mimic the effects of self-shielding
  of H$_2$ and CO, which is not explicitly included in the code, the
  photodissociation rates for both these species were set to zero in the PDR
  models. All the DC and PDR models produced in this work are also summarised
in Table~\ref{models_used}.  A range of PDR models were run to
  demonstrate the effect of different aspects of the HNCO chemistry, different
  densities and different initial chemical compositions, either atomic or
  evolved from a collapsing cloud. A very dense PDR with an initially evolved
  chemistry such as models PDR4 and PDR5 were explored to represent clumps
  exposed to the UV radiation from a very nearby young massive star.

\begin{table*}[t]
\centering

\caption{Summary of the dark cloud (DC), hot core (HC) and PDR models
  used to model the time dependent HNCO gas-phase abundance.  $T$ is
  the gas-phase temperature and $n_{\mathrm{H}}$ is the total hydrogen
  number density (both held constant).  The initial chemistry was
  either $atomic$, where all species (except H) were in their
  elemental form, or $evolved$, where abundances from a collapsed
  cloud (CC) model were used.  The collapsing cloud models are listed
  in a separate table. }

\label{models_used}
\begin{tabular}{cccll}
\hline\hline
Model & $T$~(K) & $n_{\mathrm{H}}$~(cm$^{-3}$) & Initial chemistry & HNCO formation chemistry \\
\hline
DC1 & 10 & 2.0$\times10^{4}$ & Atomic & Reaction c only\\
DC2 & 10 & 2.0$\times10^{4}$ & Atomic & Reactions c,d,e and f\\
DC3 & 10 & 2.0$\times10^{4}$ & Atomic & Reactions d,e,f and g\\
DC4 & 10 & 2.0$\times10^{4}$ & Atomic & Reactions c,d,e,f and g \\
HC1 & 200 & 2.0$\times10^{7}$ & CC1 & All\\
HC2 & 200 & 2.0$\times10^{7}$ & CC2 & Gas phase only\\
HC3 & 200 & 2.0$\times10^{7}$ & CC3 & Surface only (reactions a,b)\\
HC4 & 200 & 2.0$\times10^{7}$ & CC4 & All\\
HC5 & 20 & 2.0$\times10^{7}$ & CC5 & All\\
HC6 & 20 & 2.0$\times10^{7}$ & CC1 & All\\
HC7 & 50 & 2.0$\times10^{7}$ & CC6 & All\\
HC8 & 50 & 2.0$\times10^{7}$ & CC1 & All\\
HC9 & 100 & 2.0$\times10^{7}$ & CC7 & All\\
HC10 & 100 & 2.0$\times10^{7}$ & CC1 & All\\
PDR1 & 50 & 2.0$\times10^{4}$ & Atomic & All\\
PDR2 & 50 & 2.0$\times10^{4}$ & Atomic & Gas phase only\\
PDR3 & 50 & 2.0$\times10^{4}$ & Atomic & Surface only (reactions a,b)\\
PDR4 & 50 & 2.0$\times10^{7}$ & CC1 & All\\
PDR5 & 50 & 2.4$\times10^{5}$ & CC1 & All\\
\hline\hline
\end{tabular}
\end{table*}

\subsection{Chemical model}%need to split into 3 sentences%

For the CC and DC models, along with PDR1, PDR2 and PDR3, all species
were initially present in their elemental (atomic) form.  The only
exception was hydrogen which was assumed to be predominantly molecular
with only a small fraction present as atoms.  These initial abundances
(relative to $n_\mathrm{H}$) are given in Table \ref{abundances}.  For
the HC models, along with PDR4 and PDR5, an evolved initial chemistry
was used.  The initial abundances at the start of these models used
the abundances taken from a CC model. All the HC models and PDR4 used
abundances taken from a CC at the final density,
$2\times10^{7}$~cm$^{-3}$, whereas PDR5 used abundances from a CC
model at an earlier stage in the collapse.  In this case the density
reached in the collapse was equal $2.4\times10^{5}$~cm$^{-3}$. 

For all the models, once the chemistry had been initialised, the
subsequent evolution of all species was followed via a chemical
reaction network.  This network included both reactions in the gas
phase and on dust grain surfaces, along with the coupling mechanisms
of accretion and thermal desorption only.  
%Non-thermal desorption mechanisms were neglected as the lifetimes of such mechanisms are greater than both the collapse and freeze-out times, and once temperatures much greater than 10~K are reached thermal desorption can be considered the dominant mechanism \citep[e.g.][]{tth1999,i1977}

The gas-phase chemistry included all the reactions from the latest UMIST
database for astrochemistry (UDfA\footnote{UDfA; {\tt www.udfa.net}}), Rate06
\citep{wetal2006}.  For the grain-surface chemistry over 200 reactions were
included from \citet{ar1977}, \citet{hhl1992} and \citet{hh1993}.  Reactions
on the surface were treated using the modified rate equation method as used by
\citet{rh2000}.

In order to more completely model the HNCO
gas-phase abundance some additional chemistry was included. Five
surface reactions involving HNCO were taken from the expanded OSU
gas-grain reaction network \citep{gwh2008} comprising of a single
formation path (reaction b in Table~\ref{reactions}) plus four
destruction mechanisms. This surface chemistry was complemented by
some gas phase reactions \citep[also from][]{gwh2008} which followed
the destruction of four new species (HNCHO, HNCOCHO, HNCONH and
HNCOOH) formed from surface processing of HNCO.  The destruction of
HNCO in the gas phase was also included.  This followed the treatment
of \citet{tth1999} and included the reactions between HNCO and the
following species; H$^+$, H$_3^+$, He$^+$, C$^+$, H$_3$O$^+$, CO$^+$,
HCO$^+$, OCN$^+$ and HNCO$^+$. The rate coefficients for these
reactions were taken from the ion-neutral reaction database compiled
by \citet{anicich}. The photodestruction of HNCO by FUV photons and
cosmic ray induced photodissociation was also incorporated
\citep{getal1989,retal1991}. { Reactions with ions, particularly H$_3^+$ and He$^+$ 
dominate the destruction of HNCO in both dark clouds and hot cores.}

Two gas-phase formation reactions for
HNCO were added (reactions $c$ and $d$ in Table~\ref{reactions}). The
first reaction ($c$) is a neutral-neutral reaction whilst the second ($d$)
is the dissociative recombination of H$_2$NCO$^{+}$.  For reaction $c$
the rate coefficient is poorly known. \citet{tth1999} adopted a value
of $10^{-13}$ cm$^3$ s$^{-1}$ in their dark cloud model. However the
energy barrier for this reaction may be much larger than suggested by
this number and as such this reaction may only become significant in
regions much warmer than dark clouds. The rate coefficient for this
reaction has been measured but only at temperatures above 300~K (see
the NIST chemical kinetics database\footnote{The NIST chemical
  kinetics database;
  \tt{http://kinetics.nist.gov/kinetics/index.jsp}}).  A fit to the
available data provides an expression for the temperature dependent
rate coefficient;
\begin{equation}
\label{k_rate}
k(T) = 1.5 \times 10^{-11} \times \mathrm{exp}\left(\frac{-4465}{T}\right) \mathrm{cm^3 s^{-1},}
\end{equation}
where $T$ is the temperature.  At 10~K this gives a considerably
smaller value than that used by \citet{tth1999}.  We used model DC1 to
examine how the rate of reaction $c$ affects the HNCO abundance - the
model was run several times, each time using a different value for the
rate coefficient ranging from $10^{-22}$~cm$^3$ s$^{-1}$ to
$10^{-8}$~cm$^3$ s$^{-1}$ (increasing the value by two orders of
magnitude each time).  For all other models in which reaction $c$ was
included the rate was calculated using Equation~\ref{k_rate}.

Three further reactions were added that form `parent' species needed
for the formation of HNCO.  Reactions $e$ and $f$ are ion-neutral
reactions that are the first two reactions in a three stage process.
The actual formation of HNCO in this reaction channel is via reaction
$d$ \citep{i1977}.  The rates of these reactions are calculated using
the Langevin rate.  The final reaction ($g$) is a charge exchange
reaction between He$^+$ and OCN which produces OCN$^{+}$.  This
reaction represents a third, new, reaction channel between these two
species since the UDfA already contains two reactions between He$^+$
and OCN.  A rate coefficient of $10^{-9}$ cm$^3$ s$^{-1}$ was adopted
for reaction $g$ - representative of charge exchange reactions.

\begin{table}
\centering
\caption{Standard initial chemical abundances relative to the total hydrogen number density; $Y(\mathrm{X})=n(\mathrm{X})/n_\mathrm{H}$.}
\label{abundances}
\begin{tabular}{lclclc}
\hline\hline
 & $Y(\mathrm{X})$ &  & $Y(\mathrm{X})$
&  & $Y(\mathrm{X})$ \\
\hline
H$_{2}$ & 0.5 & H & $1.00\times 10^{-6}$ & He & 0.1 \\
C & $1.32\times 10^{-4}$ & N & $7.50\times 10^{-5}$ & O & $3.19\times 10^{-4}$\\
S & $1.86\times 10^{-5}$ & Si$^+$ & $2.46\times 10^{-7}$ & Fe$^+$ & $1.50\times 10^{-8}$ \\
\hline
\end{tabular}
\end{table}
\begin{table}
\centering
\caption{Key reactions included in the chemical network relevant to the formation of HNCO as included in the chemical network. The prefix G denotes surface species. AR77;\citet{ar1977}, GWH08;\citet{gwh2008}, TTH99; \citet{tth1999}, I77; \citet{i1977}}
\label{reactions}
\begin{tabular}{crcll}
\hline\hline
 Label & \multicolumn{3}{c}{Reaction} & Reference \\
\hline
$a$ & GOCN + GH &$\rightarrow$& GHNCO & AR77\\
$b$ & GNH + GCO &$\rightarrow$& GHNCO & GWH08\\
$c$ & OCN + H$_2$ &$\rightarrow$& HNCO + H & TTH99\\
$d$ & H$_2$NCO$^{+}$ + e$^{-}$ &$\rightarrow$& HNCO + H & I77\\
$e$ & H$_2$ + OCN$^{+}$ &$\rightarrow$& HNCO$^{+}$ + H & I77\\
$f$ & H$_2$ + HNCO$^{+}$ &$\rightarrow$& H$_2$NCO$^{+}$ + H & I77\\
$g$ & He$^{+}$ + OCN &$\rightarrow$& He + OCN$^{+}$ & - \\
\hline
\end{tabular}
\end{table}

All models were allowed to evolve chemically for 10$^7$ years, except
for the collapsing cloud models which were stopped once the final
density was reached.
%%%%%%%%%%%%%%%%%%%%%%%%%%%%%%%%%%%%%%%%%
\section{Results}
\subsection{Gas phase formation of HNCO in dark clouds}
\label{results}

\begin{figure}
  \resizebox{\hsize}{!}{\includegraphics{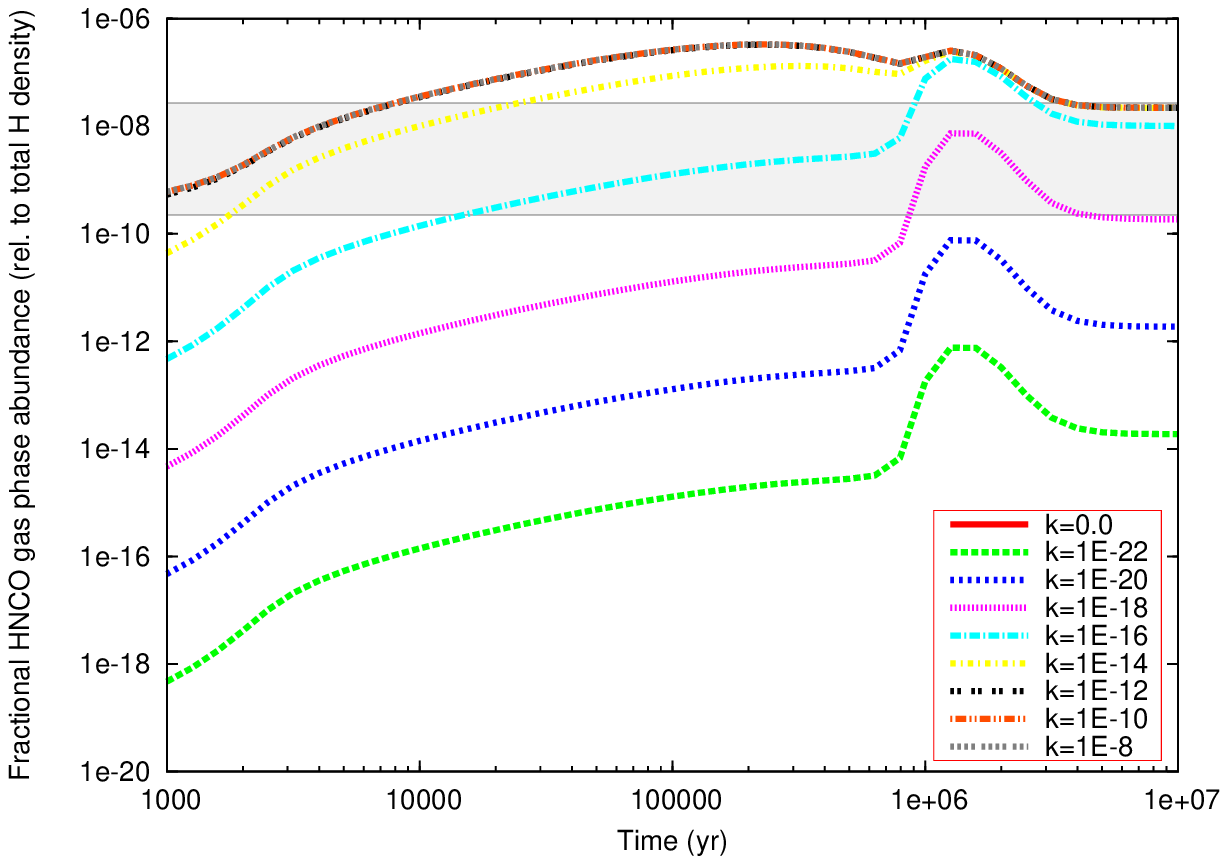}}
  \resizebox{\hsize}{!}{\includegraphics{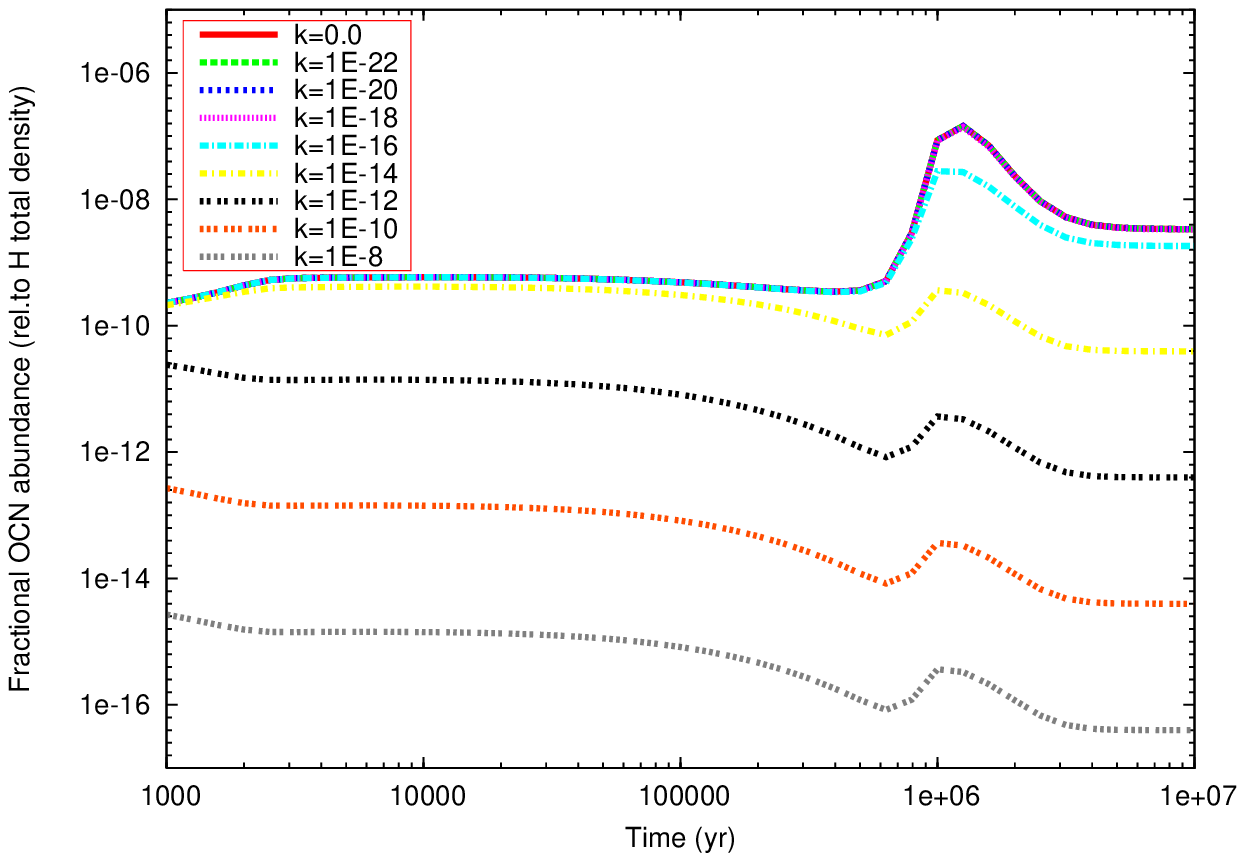}}
%  \resizebox{\hsize}{!}{\includegraphics{figures/turner_rate_ocn+.dat.eps}}
  \caption{Time dependent, gas phase relative abundances for HNCO
    (\emph{top}}) and OCN (\emph{bottom}) as a function of rate
    coefficient for reaction $c$ in a non-collapsing dark cloud (model
    DC1).  All other HNCO formation pathways are prohibited, both in
    the gas-phase and on grain surfaces. Reaction $g$ has rate
    coefficient of 0.0.  The shaded region defines the range of values
    for the observed HNCO abundance, $1.4\times10^{-10}$ to
    $5.4\times10^{-8}$.
  \label{turner1}
\end{figure}

For the first part of our investigation, using DC1, we explored the
effect the rate coefficient for reaction $c$ has on the abundance of
HNCO in the gas phase.  Figure~\ref{turner1} shows the HNCO abundance
along with that of OCN for a range assumed rate coefficients for
reaction $c$. For a rate coefficient of $10^{-18}$cm$^{3}$ s$^{-1}$
and above it is possible to achieve HNCO abundances within the
observed range of $1.4\times10^{-10}$ to $5.4\times10^{-8}$ \citep[][]{metal1996,betal2007}.
Equation~\ref{k_rate} implies that for the $\sim4500$~K barrier which
laboratory data (NIST) infer for this reaction, temperatures above
277~K are required to reach this value of the rate coefficient for
reaction $c$. Alternatively for this reaction to be effective at a gas
temperature of 10~K, the barrier would have to be less than
$\sim165$~K, much lower than measurements indicate. In other
  words, at the temperatures of dark molecular clouds these reactions
  do not occur.

% For comparison, to match the observed abundances with the rate
% coefficient of $10^{-13}$~cm$^{3}$ s$^{-1}$ used by \citet{tth1999},
% the implied reaction barrier is only approximately 70~K (calculated
% from scaling the rate for a typical neutral-neutral fast reaction with
% the Boltzmann factor).  \citet{tth1999} used this value of $k$ based
% on the measured rate for the reaction of C$_2$H$+$H$_2$ at 300~K
% \citep{h1994}; a reaction with an energy barrier of $\sim1200$~K.  To
% achieve a rate of $10^{-13}$~cm$^{3}$ s$^{-1}$ with the infer energy
% barrier of 4500~K a temperature of $\sim900$~K would be required.
% Therefore, for at the 10~K temperature of model DC1 in which HNCO is
% formed exclusively by reaction $c$, it is impossible to reproduce the
% observed $10^{-9}$ to $\sim10^{-8}$ fractional abundances of HNCO in
% dark clouds with a rate coefficient consistent with laboratory data.
% Indeed for the large barrier for reaction $c$, gas phase synthesis is
% not a significant source of HNCO even in hot cores.
 
\begin{figure}
  \resizebox{\hsize}{!}{\includegraphics{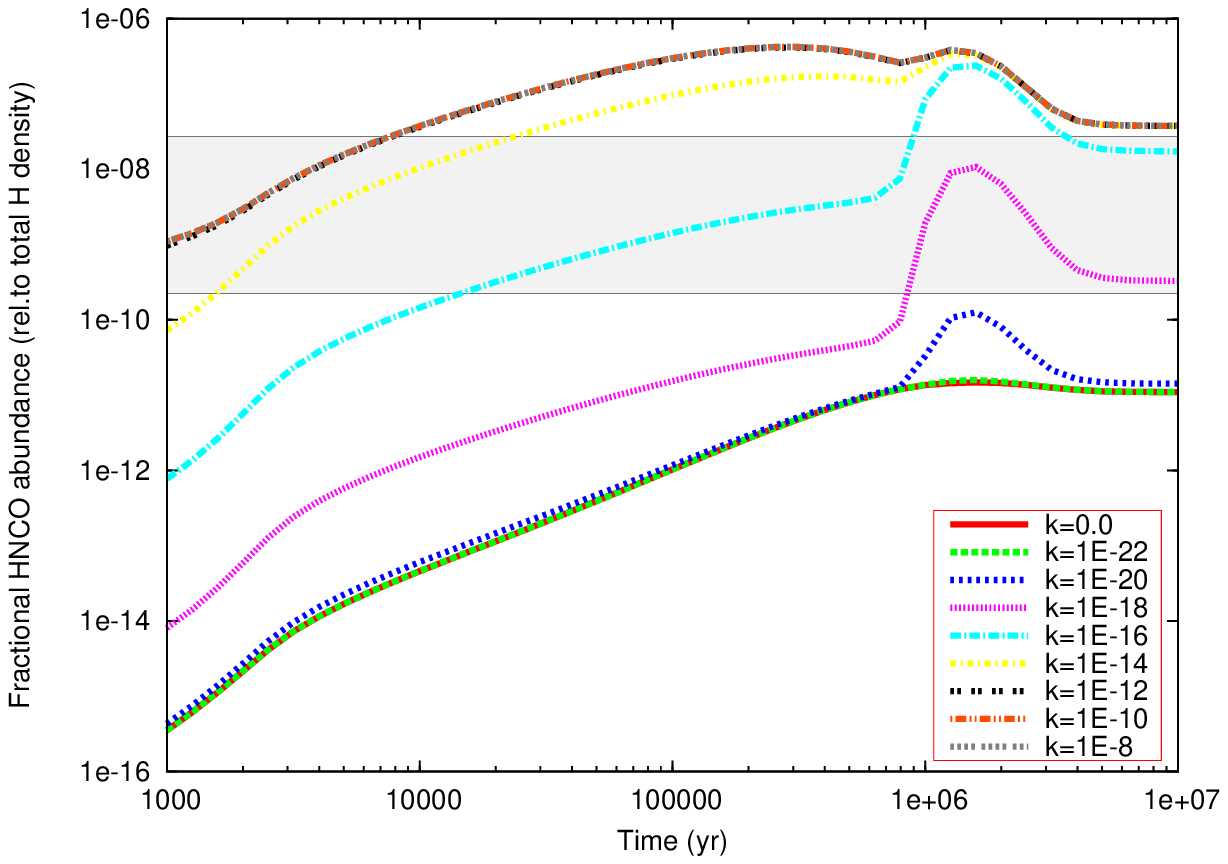}}
  \resizebox{\hsize}{!}{\includegraphics{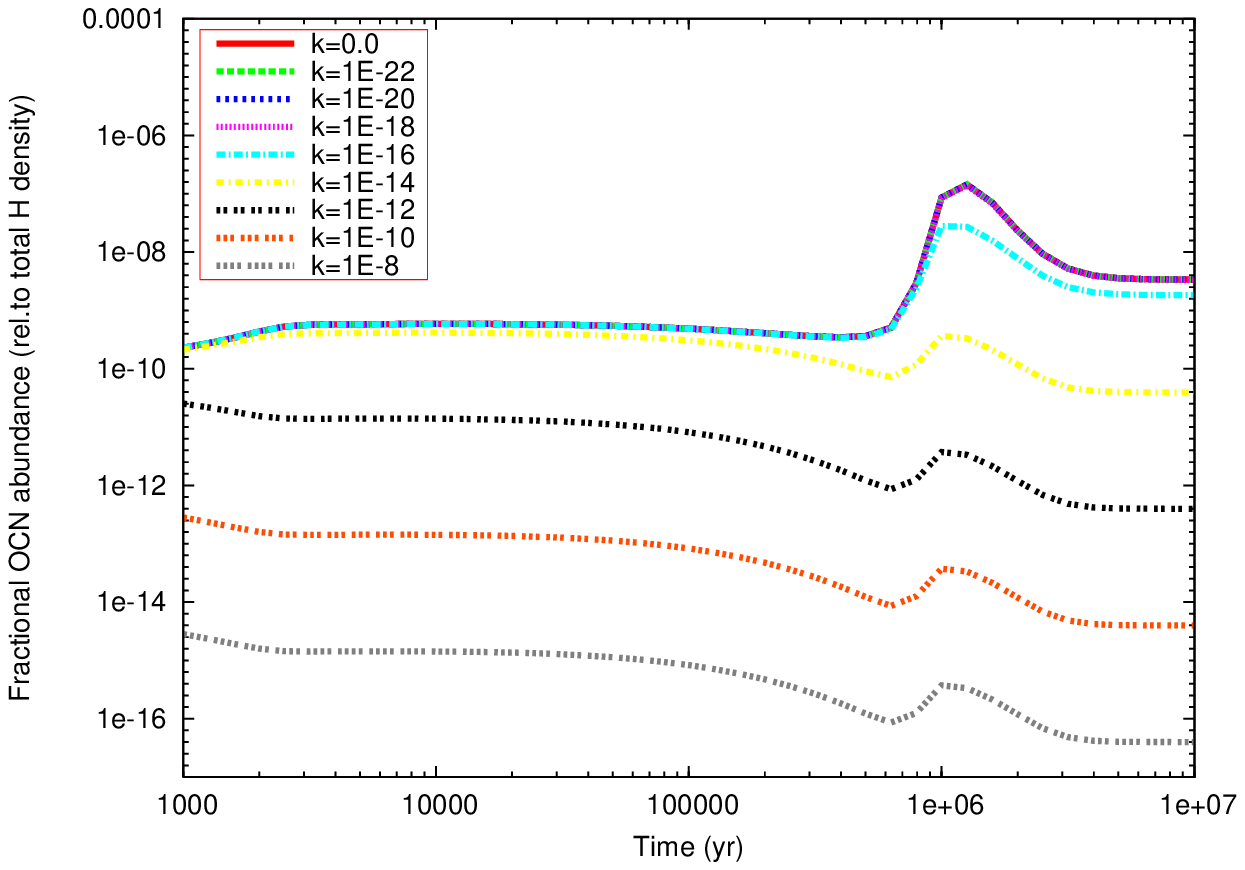}}
  \resizebox{\hsize}{!}{\includegraphics{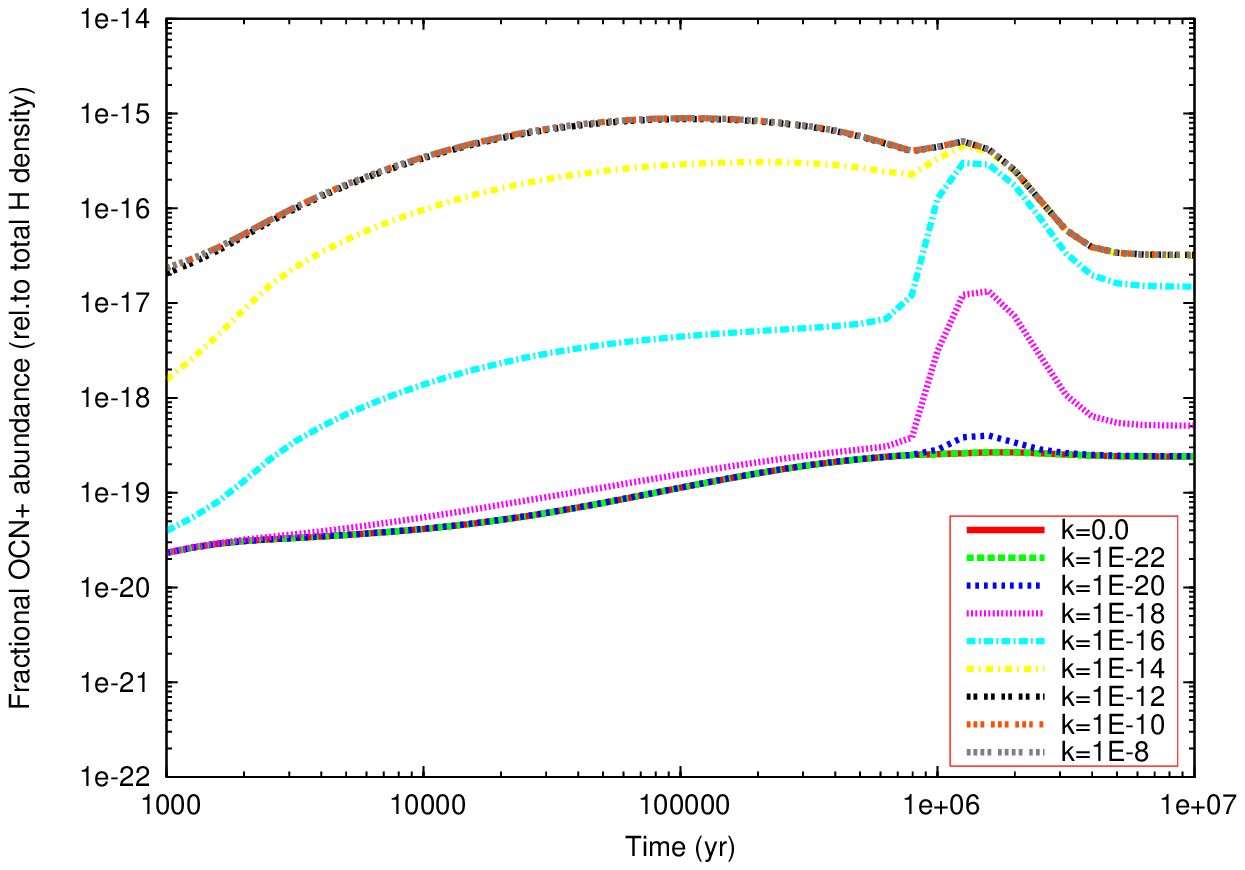}}
  \caption{Time dependent, gas phase relative abundances for HNCO
    ($top$), OCN ($middle$) and OCN$^+$ ($bottom$) as a function of
    rate coefficient for reaction $c$ in a non-collapsing dark cloud
    (model DC2).  All gas-phase HNCO formation pathways are allowed;
    grain-surface reactions are prohibited. Reaction $g$ has rate
    coefficient of 0.0. The shaded region defines the range of values
    for the observed HNCO abundance}
  \label{turner2}
\end{figure}

\begin{figure}
  \resizebox{\hsize}{!}{\includegraphics{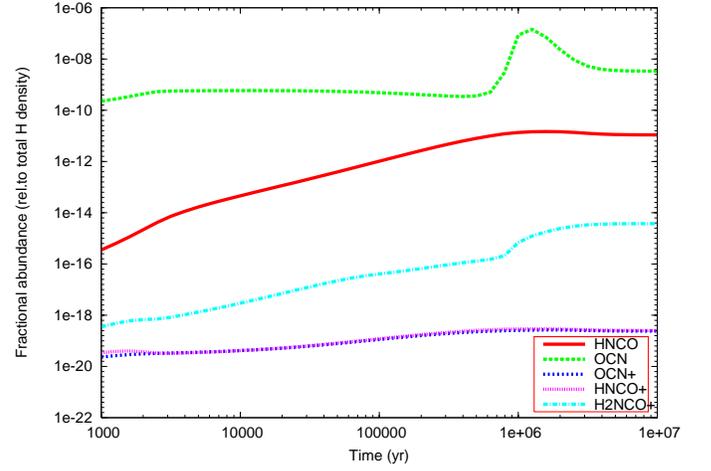}}
  \caption{Time dependent, gas phase relative abundances for various species
    in a non-collapsing dark cloud using only reactions $d$,$e$ and $f$ to
    produce HNCO (model DC3).}
  \label{iglesias}
\end{figure}

\begin{figure}
  \resizebox{\hsize}{!}{\includegraphics{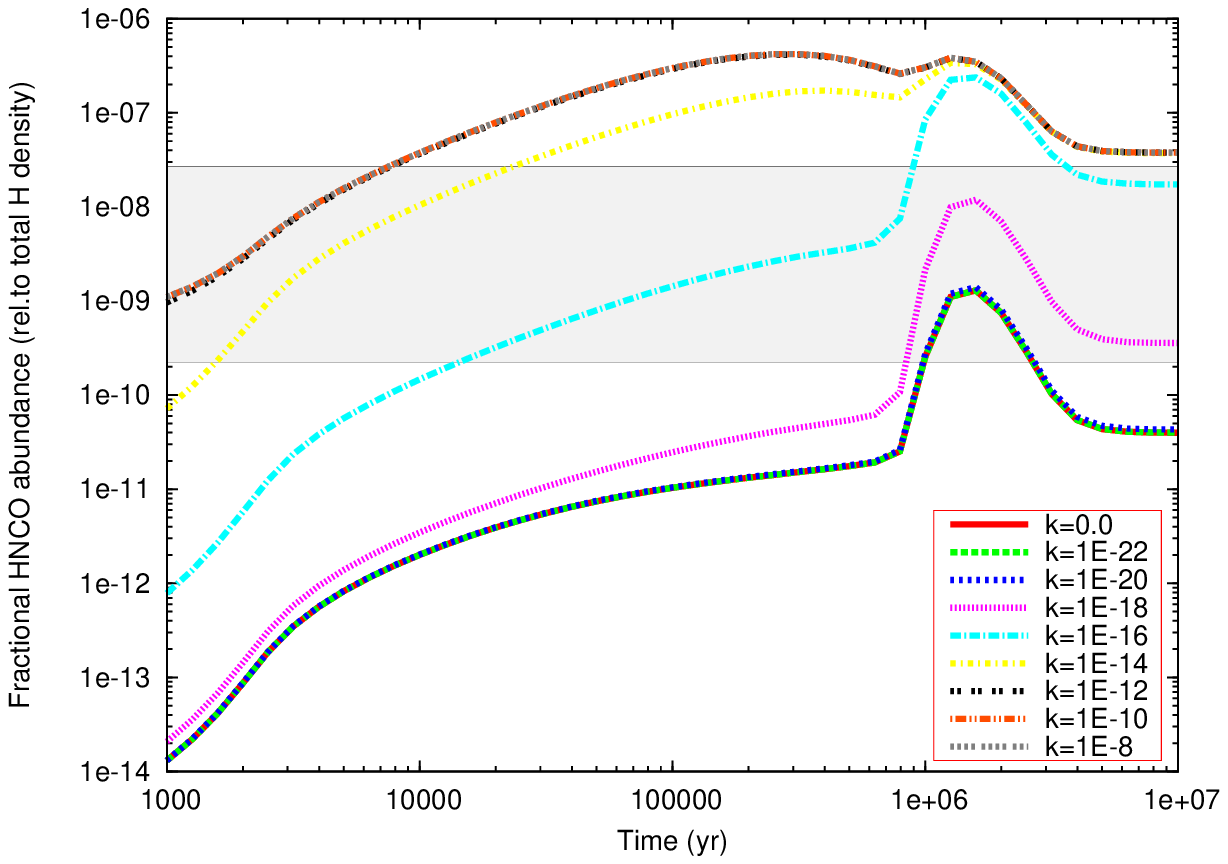}}
  \resizebox{\hsize}{!}{\includegraphics{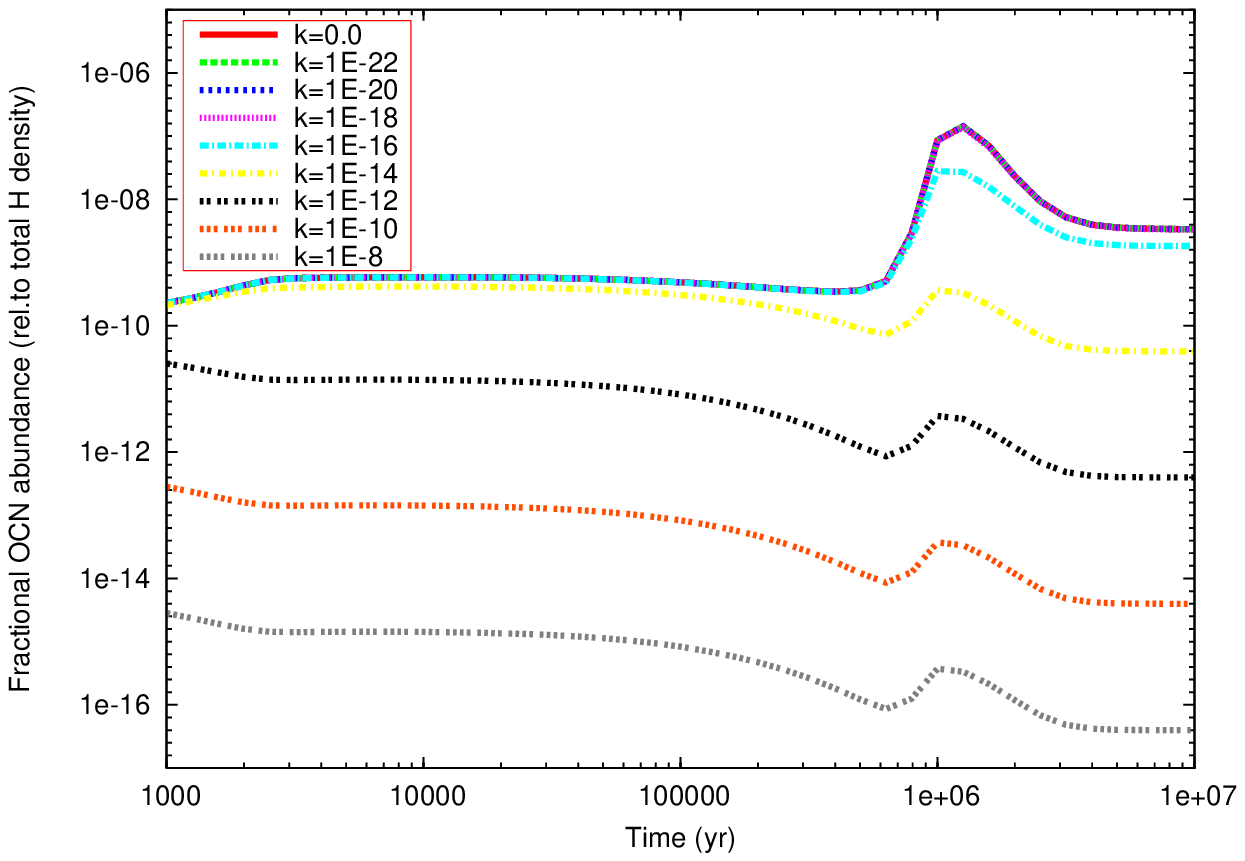}}
  \resizebox{\hsize}{!}{\includegraphics{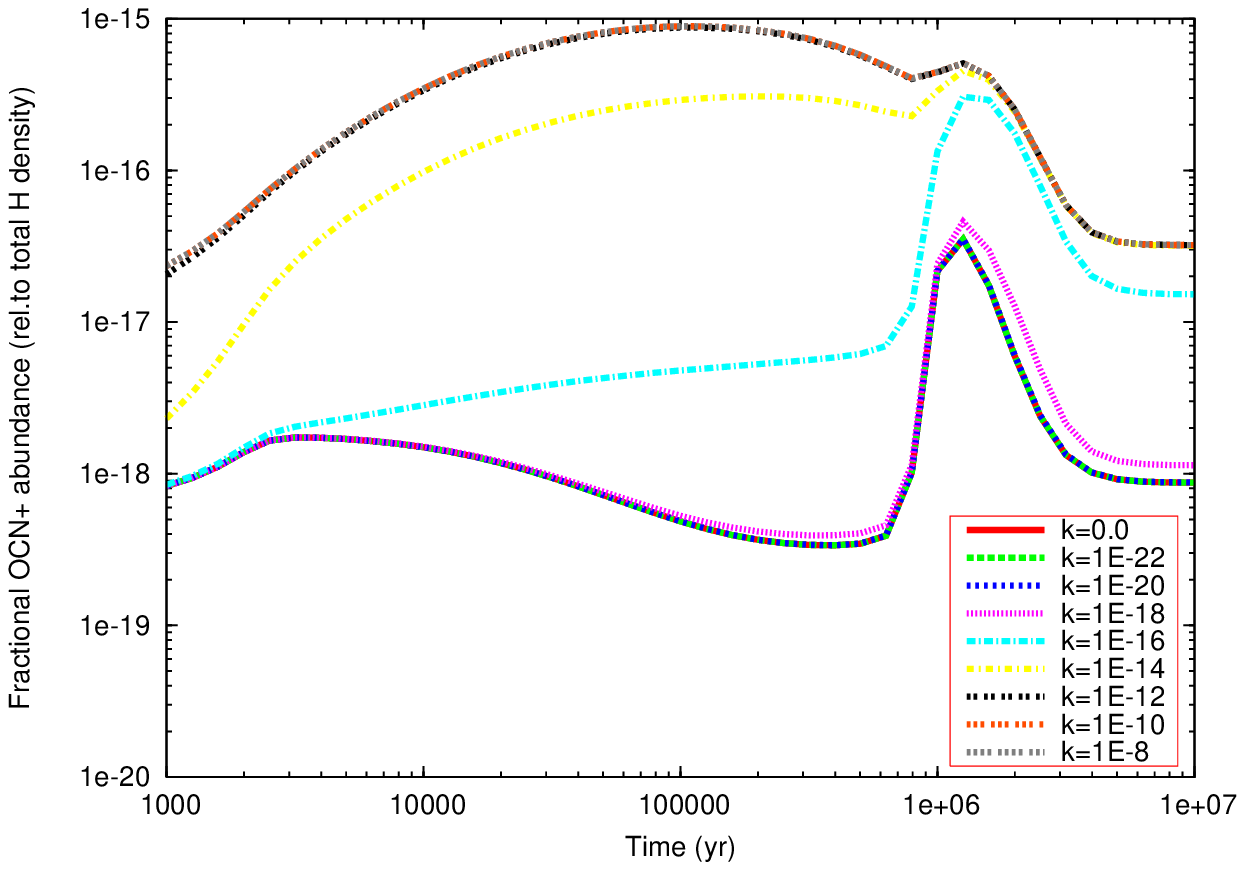}}
  \caption{Time dependent, gas phase relative abundances for HNCO ($top$), OCN
    ($middle$) and OCN$^+$ ($bottom$) as a function of rate coefficient for
    reaction c in a non-collapsing dark cloud (model DC4).  All gas-phase HNCO
    formation pathways are allowed but grain-surface reactions are
    prohibited. Reaction $g$ has rate coefficient of 1.0$\times10^{-9}$ cm$^{3}$
    s$^{-1}$. The shaded region defines the range of values for the observed
    HNCO abundance}
  \label{turner3}
\end{figure}

Model DC2 illustrates how the other gas-phase formation route (via
reaction $d$) influences the HNCO abundance.  Reaction $c$ was also
included in this model and the rate was varied as before.  Figure
\ref{turner2} shows the HNCO and OCN abundances along with those for
OCN$^+$.  A lower limit for the fractional HNCO abundance (at steady
state) is effectively set at 1.1$\times10^{-11}$.  This corresponds to
the DC2 model with a rate cofficient of 0 for reaction $c$ (the solid
red line in Figure~\ref{turner2}).  In this model HNCO is
underabundant compared to observations by over 2 orders of magnitude.
However, as seen for DC1, when the rate coefficient for reaction $c$
increases, it is possible to achieve higher abundances.  There is
however very little difference between DC1 and DC2 for those models
with non-zero rate coefficients for reaction $c$.  Figure \ref{iglesias}
shows the gas-phase abundance for the key reactants/products involved
in the gas-phase production of HNCO (model DC3).  This shows that
there is OCN in the gas-phase but due to the low temperature
this is not converted into HNCO.  Instead, at low temperatures HNCO
formation through reaction $d$ (via reactions $e$ and $f$) clearly
dominates.
The fourth and final DC model, DC4, includes all the gas phase
formation reactions listed in Table \ref{reactions} (but in
  common with the other DC models, no grain
  surface processes) and the reaction rate for reaction $c$ has been
varied.  The abundance profiles for this model are seen in Figure
\ref{turner3}.  For small values of the rate coefficient of reaction $c$, 
the inclusion of reaction $g$ increases the HNCO abundance. However, to achieve 
HNCO abundances in the observable range for
  ages less than $10^6$ years, the Turner rate must be greater than
$10^{-16}$ ~cm$^3$ s$^{-1}$, which would imply temperatures of
$>370$~K.  At ages greater than $\sim10^6$ years reaction $g$
  appears to be able to produce sufficient HNCO to match the
  observations even for a much reduced rate for reaction $c$ (indeed
  even a rate of zero). However interpreting this late time rise in
  HNCO abundance needs to be treated with caution as these models only
  include gas phase reactions and so do not include the effect on the
  gas phase abundance of freeze-out on to the grains which will be
  very significant at these late times.  This peak in HNCO abundance
  might occur earlier in models of denser clouds, but the effect of
  freeze-out on the gas phase abundances would also occur earlier.


\subsection{HNCO in photodissociation regions}

\begin{figure}
  \resizebox{\hsize}{!}{\includegraphics{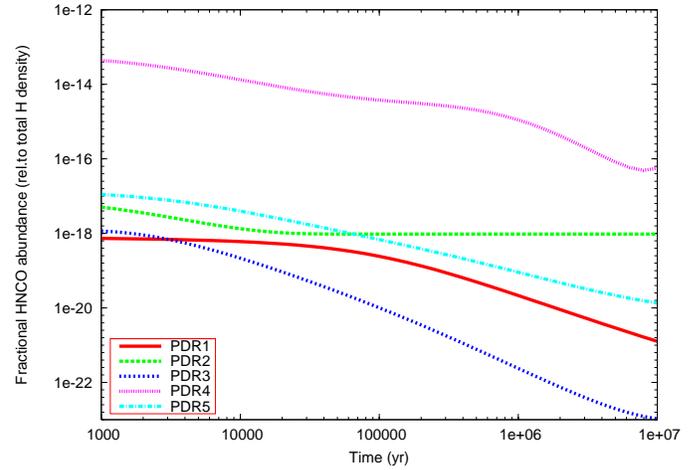}}
  \caption{Time dependent, relative gas phase abundance of HNCO for
    the five PDR models. { For comparison the observed value is between
      $1.4\times10^{-10}$ abd $5.4\times10^{-9}$. } Note that curves
    for PDR1 and PDR2 are coincident. }
  \label{pdrmodels}
\end{figure}

Figure \ref{pdrmodels} shows the results for our PDR models.  The HNCO
abundance in PDR regions are even lower than those found in DC.  For the three
PDR models with an initially atomic chemistry (PDR1, PDR2 and PDR3) the
largest late time abundance of HNCO ($10^{-17}$) is produced in the models
with gas phase formation (PDR1 and PDR2), but these produce up to 9 orders of
magnitude less HNCO than in a DC model.  In model PDR3, which allows only surface
reactions, HNCO is even less abundant, reaching a maximum value of $10^{-20}$.
The final two PDR models are also presented in Figure~\ref{pdrmodels}.  Both
models follow a similar trend, that is the HNCO abundance is steadily
decreasing with time, reaching a plateau at later times.  However, the HNCO
abundance in PDR4 is initially larger by about two orders of magnitude than in
PDR5, although they ultimately reach similar abundances of about
$2\times10^{-18}$ .
None of the PDR models ever approach the abundances of HNCO seen in DCs and
HCs - even for the PDR models with start with previously evolved chemistries
(PDR4 and PDR5).

The warmer conditions in these PDR models, largely due to the reduced optical depth
and enhanced radiation field resulting in the heating of dust grains, lead to faster desorption meaning that once
formed HNCO is never on the grains for long (at the 50~K used in these models
the surface lifetime is approximately 100 years) so surface formation is
limited.  Furthermore, HNCO is efficiently destroyed by both FUV and cosmic
ray induced photodestruction reactions.  At no time during the evolution of
any of our PDR models does the HNCO abundance come close to the $\sim10^{-9}$
observed by \citet{metal2008} towards PDR regions.  As a consequence of this
low HNCO abundance, the ratio of the abundance of HNCO to the CS abundance in
these PDR models is also many orders of magnitude lower than seen by
\citet{metal2008} (Fig.~\ref{pdr_cs}).  The origin of the HNCO observed by
\citet{metal2008} apparently in PDRs is therefore unclear.
% These discrepancies suggest that perhaps the HNCO emission observed by
% \citet{metal200} comes from more embedded (shielded) material in a PDR. More
% detailed PDR models than the simplistic ones produced here are needed to
% explore this possibility.

\subsection{HNCO in hot cores} 

\begin{figure}
  \resizebox{\hsize}{!}{\includegraphics{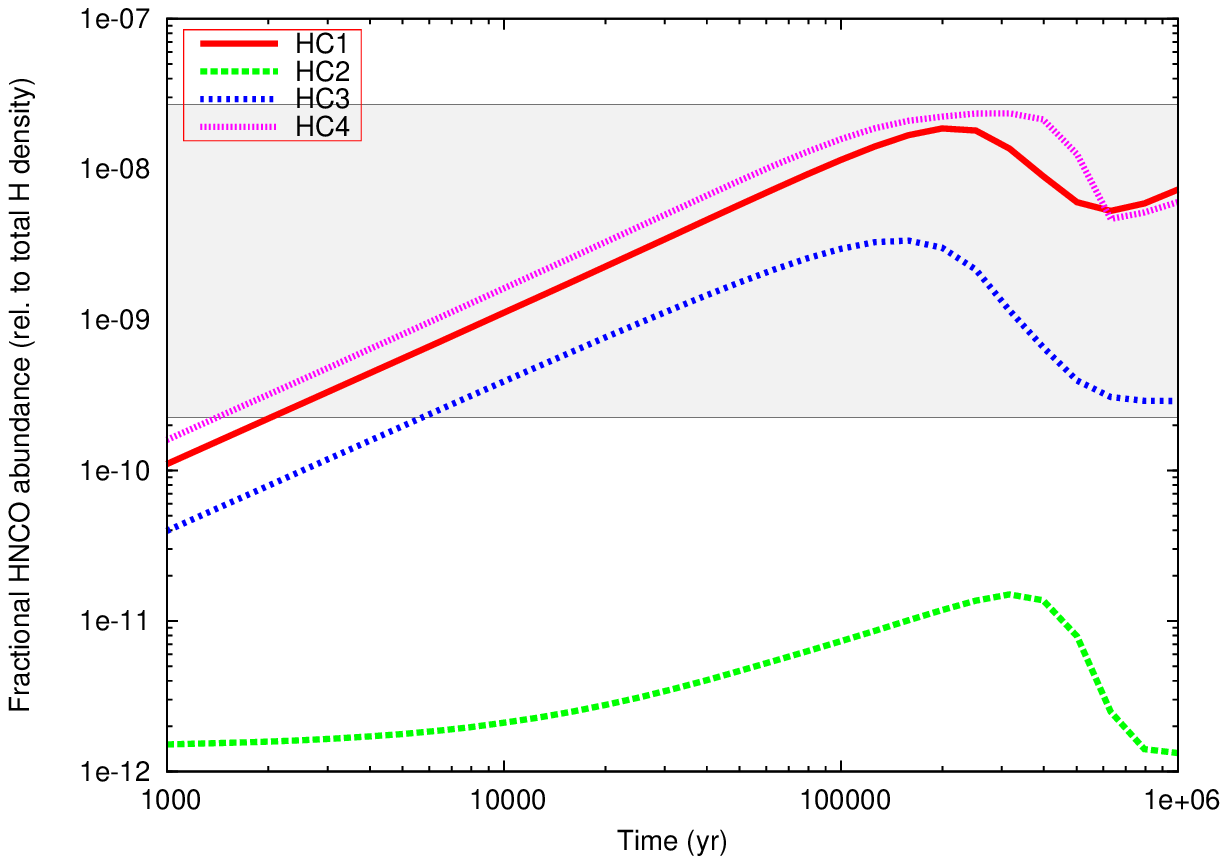}}
  \resizebox{\hsize}{!}{\includegraphics{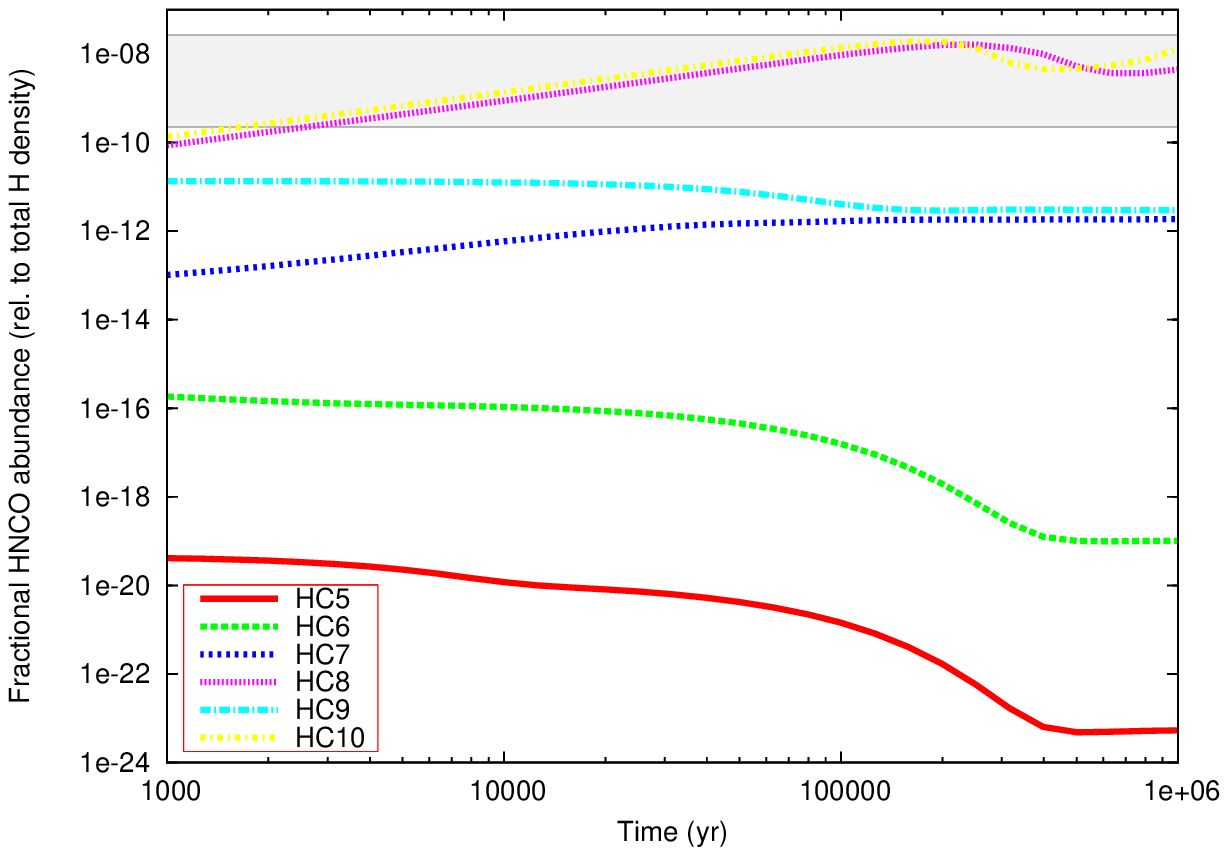}}
  \caption{Relative HNCO gas-phase abundance as a function of time for hot core models (HC) 1 - 4 ($top$) and 5 - 10 ($bottom$).}
  \label{hc1}
\end{figure}

Figure \ref{hc1} shows the HNCO gas-phase abundance as a function of
time for all the hot core models 1 to 10.  Only times between 1000
years to 1 Myr (since the HC's switch-on) are plotted as HCs are
transient objects and are relatively short-lived phases of high-mass
star formation lasting less than 1 Myrs.

HC1 includes the complete reaction network described earlier.  The
HNCO abundance for this model does reached observed levels.  Beyond
$\sim$~10$^4$ yrs values greater than $10^{-9}$ are seen and a peak of
$\sim$~2$\times10^{-8}$ is reached at around 0.2-0.3 Myr.  For the
model with surface formation only (HC3) the lower observed limit for
HNCO is reached but only beyond $10^5$~yr. The HNCO abundance in HC4 
(which differs from HC1 only in the elemental sulphur abundance) is almost identical in
behaviour as HC1.  The major differences are that higher abundances
are reached at slightly earlier times and the peak value of just over
2$\times10^{-8}$ remains for a further 0.1 Myr.

The gas-phase formation only HC model (HC2) shows that the abundance
for HNCO remains relatively flat and never exceeds a few $10^{-11}$,
and even then only at HC ages greater than 0.3 Myr. As discussed
above, this indicates that even the relatively high temperatures in
hot cores, the gas-phase reactions by themselves are incapable of
producing HNCO at observed levels.

The remaining HC models all had the same chemical networks (the full network with all reactions `on'), it was only
the initial and final temperatures that differed.  In fact for the sake of
comparison three of the models, HC5, HC7 and HC9, each had a constant
temperature during the initial collapse and throughout subsequent constant density
evolution.  The first two models, HC5 and HC6, had the same post-collapse
temperature (20~K), however HC6 used evolved abundances from a cloud collapsed
at 10~K wheras HC5 used abundances from a cloud at 20~K.  HC5 never has a fractional HNCO
abundance above $10^{-19}$, whilst HC6 never really exceeds $10^{-16}$.  HC6
has consistently more HNCO throughout - due to a lower temperature collapse
phase temperature.  At lower temperatures the rate of thermal desoprtion is
reduced such that the surface chemistry becomes more efficient.  This means
there is more NH, CO and OCN on grain surfaces - the key reactants for the
surface formation of HNCO. { For the models presented in this paper, the surface reaction between OCN and H is more important for the formation of HNCO than the reaction between CO and NH.}

The next two models had post-collapse temperatures of 50~K.  Again,
the only difference being the collapse temperature (HC7 with $T=50$~K,
HC8 with $T=10$~K).  Both of these models have higher HNCO abundances
compared to HC5 and HC6.  Model HC8 is almost exactly the same as HC1,
but HC7 has a relatively constant fractional abundance of around
$10^{-12}$.  Model HC9 had a collapse and post-collapse temperature of
100~K.  In this model the abundance of HNCO is roughly constant.  At
early times abundances just greater than $10^{-11}$ are seen, falling
slowly to just above $10^{-12}$ beyond 0.1 Myr.  The final model,
HC10, also has a post-collapse temperature of 100~K but the collapse
temperature is 10~K.  This has an almost identical abundance profile
as HC8. These results indicate that in order to achieve HNCO gas-phase abundances similar to those observed,
  post-collapse temperatures greater than 50~K are required.

\subsection{The abundance of CS}

Our models can also be used to investigate the abundance of CS as
function of time in hot cores (and PDRs). A detailed discussion of the
important pathways for the evolution of sulphur-bearing species,
including CS, is given in \citet{mh1990}.  As mentioned above, CS is
often used as a tracer of dense gas associated with star formation and
\citet{metal2008} suggested the use of the ratio of the abundance of
HNCO to that of CS as measure of the importance of UV radiation in a
region.

\begin{figure}
\resizebox{\hsize}{!}{\includegraphics{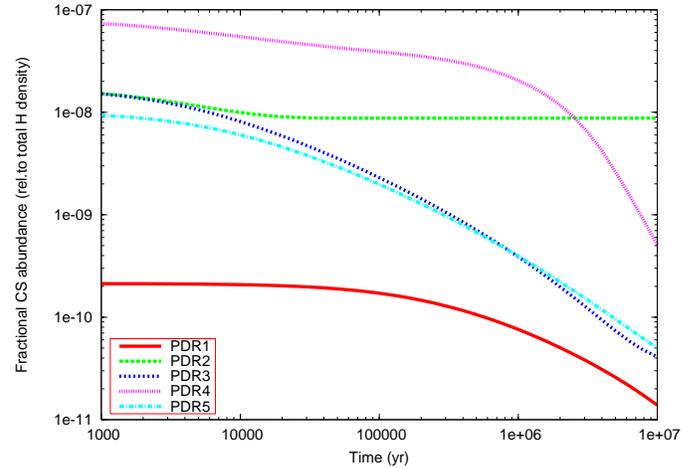}}
\caption{Time dependent, relative gas phase abundance of CS for the five PDR models.}
\label{pdr_cs}
\end{figure}

\begin{figure}
\resizebox{\hsize}{!}{\includegraphics{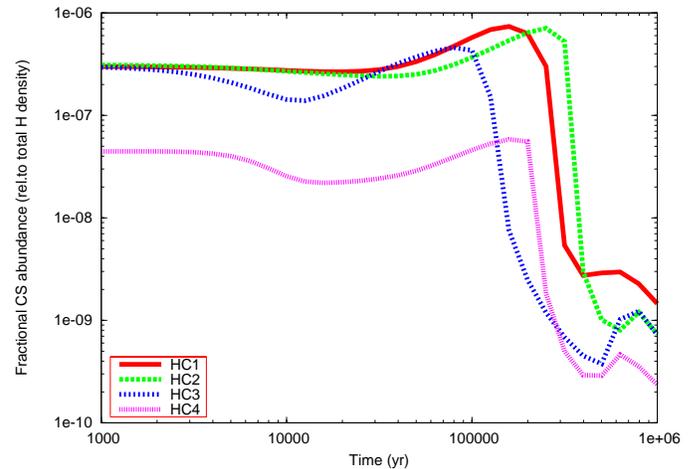}}
\caption{Time dependent, relative gas phase abundance of CS for four hot core models (HC1-4).}
\label{hc_cs}
\end{figure}

Figures~\ref{pdr_cs} and \ref{hc_cs} show the evolution of the abundance of CS
in our PDR and HC models respectively.  All three PDR models which start with
atomic compositions (PDR1, PDR2 and PDR3) show identical behaviour.  From its
initial value of $2\times10^{-10}$ the abundance of CS monotomically decreases
by a factor of $>10$ over the period covered by the model.  For PDR4 and PDR5
which start with evolved chemistries the CS abundance also monotomically
decreases from a peak at early times, becoming approximately constant at later
times. The highest abundance of CS in these models occurs are very early times
in PDR5, where the peak abundance of $\sim10^{-9}$.  In the HC models the CS
abundance is initially relatively constant, rising to a peak after about
$10^5$ yr before rapidly declining thereafter. Comparing models HC1 and HC4,
which has a sulphur abundance ten times smaller that the other models, shows
that the CS abundance throughout most the evolution of the HC scales nearly
directly with the initial sulphur abundance.

%%%%%%%%%%%%%%%%%%%%%%%%%%%%%%%%%%%%%%%%%%%%%%
\section{Discussion}
%\subsection{Grain surface formation vs. gas phase formation}

Hot core models HC1, HC2 and HC3 demonstrate that gas phase reactions alone
can not produce a high enough abundance of HNCO to match observations. Reactions on the surfaces of grains are required.  Of the models with
a complete HNCO chemistry only models HC1, 8 and 10 achieve observed HNCO
abundances. (Model HC4 can also match the observed abundance, but only differs
from HC1 in its initial sulphur adundance.) These models all have a low
collapse temperature combined with high post-collapse (hot core) temperatures.
It appears that during the collapse phase cooler temperatures are essential.
(A HC model identical to HC4, but with an inital collapse phase at 20~K,
  rather than 10~K, produces a peak HNCO abundance of only
  $\sim3\times10^{-14}$.)  1000 years into the HC phase, reaching a peak
  abundance at $\sim10^5$ years of only $\sim10^{-11}$.  The cold collapse
allows accretion on to the surface of grains and hence surface production of
HNCO.  However, during the hot core phase, high temperatures are needed to
allow the desorption of the surface species back into the gas-phase.  The
models with constant temperatures across both stages 1 and 2 of the hot core
model are seemingly incapable of producing HNCO at observed levels.
%
% GAF: I don't see how this next statement followed from the previous. Just
% drop it.
%
%  This further
% suggests that HNCO is formed dominantly through surface reactions.

Interestingly, although grain surface reactions are clearly the ultimate
source of the HNCO, the HNCO abundance does not follow the expected time
dependent behaviour of a typical parent species ejected from grain mantles.
In the log-log figures presented here the relative HNCO abundance is seen to
linearly increase with time (the actual abundance is increasing exponentially
with time), reaching a peak value of $\sim10^{-8}$, consistent with the peak
observed values, after about 2$\times10^{5}$ years (Fig.~\ref{hc1}). This
growth suggests that the gas phase HNCO is in fact a daughter product from the
species ejected from the grain mantles.

Examining the grain mantle abundances shows that although it is efficiently
formed, HNCO is also rapidly processed to more complex species.  It is the
destruction of these species after their ejection which produces the gas phase
HNCO.  In our models the HNCO is allowed to form several daughter products
(HNCHO, HNCOCHO, HNCONH and HNCOOH) on grain surfaces during the collapse
phase.  Once the hot core stage is reached it is these species that are
ejected and subsequently destroyed in the gas-phase thus returning HNCO. Such
a possibility has been previously suggested by \citet{zhm2000}, although these
authors prefered the explanation that the HNCO they observed was formed in the
gas phase in post-shock gas.  If the HNCO in our models was not destroyed on
the grains it would be returned to the gas-phase as soon as the temperature is
increased to 200~K with a relative abundance of $\sim2\times10^{-5}$.  This
abundance is several orders of magnitude overabundant compared to
observations.

Based on its excitation temperature \citet{betal2007} argue that HNCO is a
`first generation' species ejected from grains. Our models indicate that this
is not the case and suggest that this association of a molecular species with
physical conditions on the basis of their excitation temperature alone can be
misleading.

\subsection{HNCO/CS abundance ratio}

\begin{figure}
  \resizebox{\hsize}{!}{\includegraphics{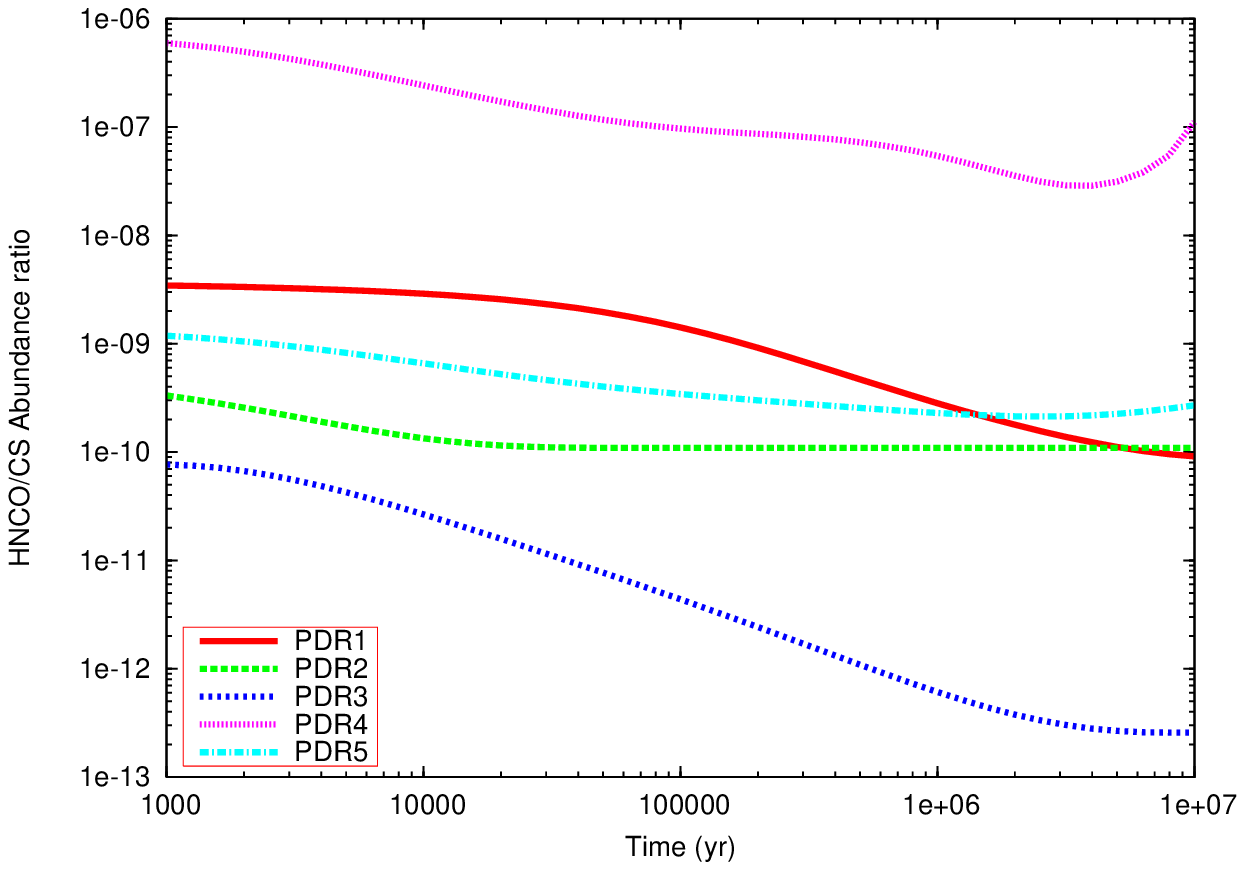}}
  \resizebox{\hsize}{!}{\includegraphics{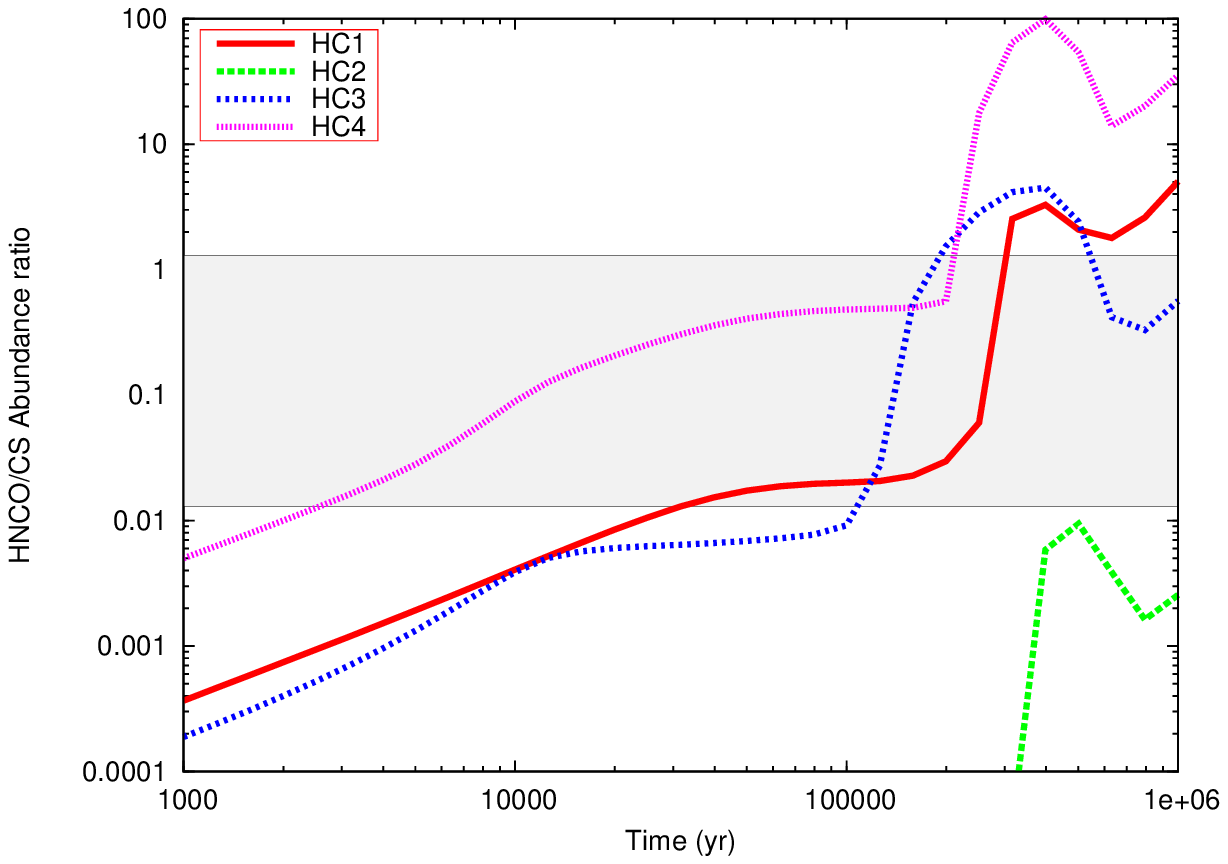}}
  \resizebox{\hsize}{!}{\includegraphics{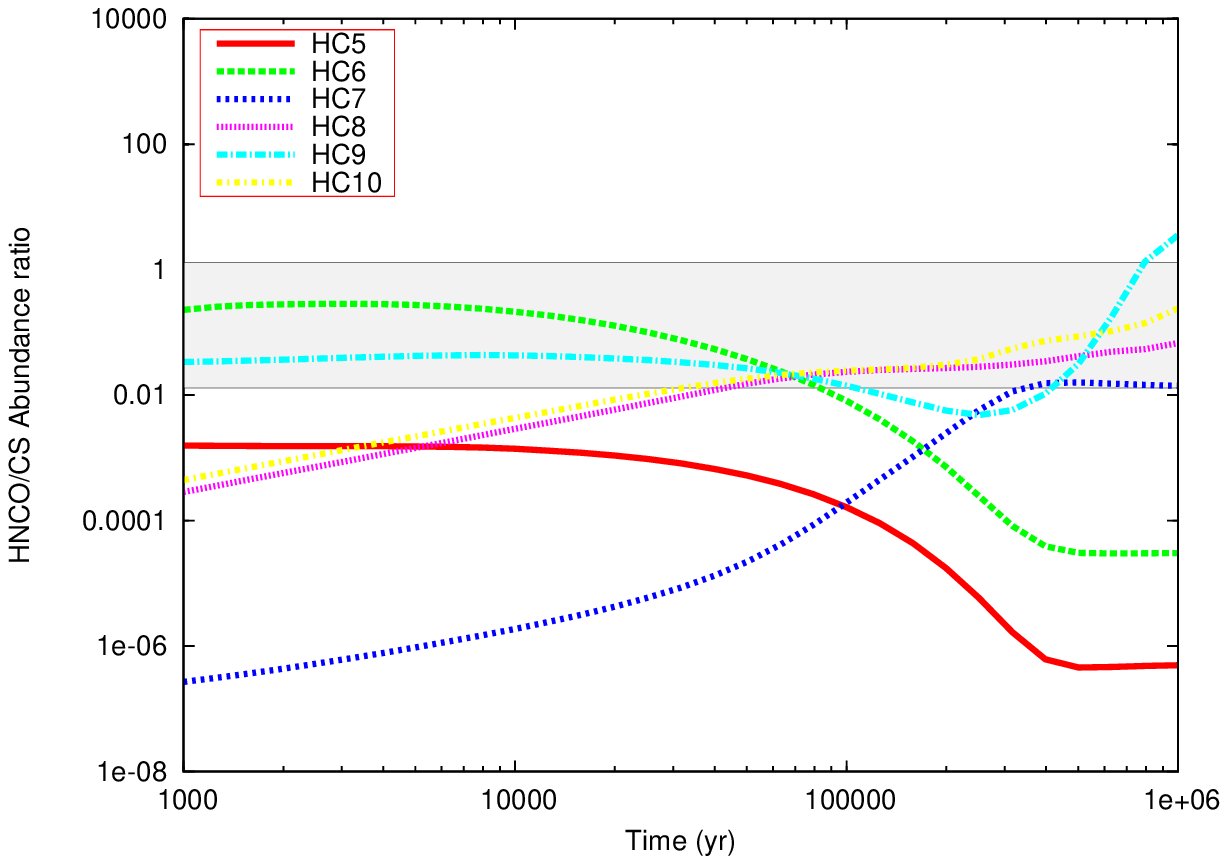}}
  \caption{HNCO/CS abundance ratio as a function of time for PDR models ($top$), hot core models 1 - 4 ($middle$) and hot core models 5 - 10 ($bottom$). The shaded region defines the upper and lower limit for the HNCO/CS ratio as found by \citet{metal2008}.}
  \label{hnco_cs}
\end{figure}

The HNCO/CS abundance ratio as a function of time is shown for the PDR models
along with all the HC models in Figure \ref{hnco_cs}.  The PDR models never
reach values greater of the HNCO/CS abundance ratio $>10^{-6}$, orders of
magnitude smaller than observed.  For model HC1 the ratio increases from about
$2\times10^{-4}$ at $10^3$ years, peaking at $\sim3$ around $3\times10^5$
years. \citet{metal2008} observed ratios for HNCO/$^{13}$CS in the range
$\sim1$ to $\sim100$. Adopting a value of 77 for the abundance ratio to
C/$^{13}$C \citep{wr1994}, this would imply HNCO/CS ratios of $\sim0.013$ to
$\sim1.3$.  The models span this ratio for times between approximately 10$^4$
years to $2\times10^5$ years showing that the time dependent evolution of the
chemistry of hot cores { alone} can produce the entire range of observed HNCO/CS
abundance ratios. { Therefore, independent of its effect on the ratio, it is 
clear that HNCO/CS is not a unique tracer of UV radiation.}

As the figure also shows, the underlaying sulphur abundance also has a direct
effect on the absolute value of the HNCO to CS abundance ratio.  When S is
reduced by a factor of 10, for a given time less than about few times $10^5$
years, the HNCO/CS ratio increases by a factor of 10. However for these times
the evolution for different sulphur abundances is essentially independent of
abundance. Overall the ratio monotonically increases with time up to a peak at
a few times $10^5$ years.  This generally monotonic behaviour suggests that
for a given sulphur abundance this ratio could act as a `chemical clock' to
constrain the ages of sources.

\section{Summary}
\label{summary}

Our model shows that gas phase reactions appear incapable of producing HNCO in
sufficient quantities (even at the $\sim200$K temperature of hot cores) to
match the observed abundance of this species.  However the rate of a key gas
phase reaction is uncertain and new laboratory determination of this rate
coefficient is highly desirable.  HNCO can be formed on icy surfaces by a
variety of reactions.  Despite the existence of these plausible mechanisms
which can produce sufficient HNCO to match observations, laboratory data is
again missing and should be obtained. Although it is produced on the grain
surfaces in potentially large amounts, our models show that the gas phase HNCO
is in fact a daughter product from the breakdown of more complex species to
which the HNCO on the grains have been processed. This is contrary to recent
speculation about the origin of this species \citep{betal2007}.  The model
also shows that in PDRs HNCO can not be directly produced in sufficient
quantities to match observations.  In hot cores in particular it seems
  that surface chemistry during an earlier cold collapse phase followed by
  gas-phase processing (in the post-collapse regime) is important in the
  production of HNCO.

The ratio of HNCO to CS evolves strongly
as a function of time while its absolute value is a sensitive function of the
sulphur abundance.  Whilst observations have suggested this ratio can be used
to trace different physical conditions, our findings show that metallicity,
choice of chemical network and hot core age can all also contribute to this
value.  Indeed, considering that HNCO contains three heavy elements one
may expect the abundance of HNCO to be highly sensitive to the metallicity of a
region. 

\begin{acknowledgements}

  DMT wishes to acknowledge the receipt of an STFC studentship. Astrophysics
  at the JBCA and QUB is supported by grants from the STFC.  This work has
  also been supported by the European Community’s human potential programme
  under contract MCRTNCT-512302 (“The Molecular Universe”). The authors would like to thank Paul Woods for his excellent proof reading ability. 

\end{acknowledgements}
% bibliography section - make sure aa.bst is either in this directory or in the texmf/bibtex/bst dir.
\bibliographystyle{aa}
% create a suitable .bib file containing all the lovely references that you have used.
\bibliography{Refs/hnco}
\end{document}